# Highly Compact Arc Bend Mode Converters for Photonic Integrated Circuits

Tushar Gaur[1] Pragya Mishra[1] and Talabattula Srinivas[1]

[1]Department of Electrical Communication Engineering, Indian Institute of Science, Bangalore – 560012

tushargaur@iisc.ac.in

**Abstract:** With the advances in on-chip photonic integrated circuits, compact and wide bandwidth mode converters are required for mode division multiplexing - demultiplexing and various other applications. We propose the design and theory for a highly compact, easy to fabricate mode converter that converts $TE_0$ to $TE_1$ and vice versa. The proposed design uses an arc bend waveguide structure based on Silicon on Insulator (SOI) platform and offers high mode conversion efficiency, low insertion loss, and large bandwidth of operation. We use coupled-mode theory to analyse the mode conversion within the structure, and FDTD simulations have been performed to verify the results. The mode conversion efficiency of 98.65% for $TE_0$ to $TE_1$ and 99.1% for $TE_1$ to $TE_0$ mode conversion with a device footprint of 10.5 μm × 2 μm is reported in this work.

## 1  Introduction:

Currently, we are witnessing a global expansion in data traffic driven by various factors. With such an outburst in the amount, processing needs to be done at high speed with large bandwidths. Silicon photonics has provided a highly competitive platform to manage and process data at such a high data rate. Silicon on insulator (SOI) offers an excellent platform for fabricating the on-chip devices for this purpose. High refractive index contrast, tight mode confinement, fabrication of compact, densely packed integrated chips, and excellent compatibility with the full form of CMOS devices are some advantages of the SOI platform. SOI is widely used in on-chip mode converters, an essential device used in the optical communication and information processing industry. Mode converters have gained interest from various research groups and industries across the globe in the past few years. Their application in mode division multiplexing-demultiplexing is the fundamental reason for the heed in this area.

With time, the larger and larger volume of information is processed on-chip in the highly-dense photonic integrated circuits, leading to the need for broadband on-chip mode converters in the last decade. Due to this reason, several research groups are working in this field [1]-[8]. To enhance the competitive capabilities of Silicon photonics against the existing electronic processors (Field programmable gate array circuits), recently paradigm shift towards the Programmable Photonic Integrated Circuits (Field-programmable photonic gate array circuits) has been observed in the optical society [9]. These are mainly made out of highly compact and densely fabricated waveguide meshes. Various functionalities, including filters, modulators, unitary operations etc., have been demonstrated on these circuits. However, to implement complete processing over data, these circuits need a compatible, compact and broadband solution for implementing mode converters in these circuits.

The present available typical mode converters studied and demonstrated are based on photonic crystals [2], dielectric met surface etched onto silicon waveguide structures [3],[7], tapered and slotted waveguide structures [4]-[6], and grating couplers [12]. The major disadvantage associated with these structures is their bandwidth of operation [4]. Most of these devices offer an operational bandwidth between 10 nm to 60 nm. Another drawback associated with some of these structures is their larger footprint, especially when operating bandwidth increases. Few mode converters using electro-optic and anisotropic materials with dielectric metasurface etched onto them were reported recently [8], but the complexity of fabrication increases as we introduce metasurface or other nanoscale periodic structures [1].

Bend waveguides have played a crucial role in today's integrated photonics, whether it is for interconnecting two devices on application-specific or waveguide meshes based photonic integrated circuits (PIC) or being used as a sensor for different applications such as biochemical sensing, temperature sensing, refractive index sensing etc., [10],[11]. The present article focuses on the telecommunication bandwidths of 1.5 μm to 1.6 μm, and we propose compact, wide bandwidth, low insertion loss and easy to fabricate waveguide mode converters based on the bent waveguide that offers a high conversion efficiency. We have used the coupling between the straight waveguide modes caused by the index perturbation due to a bend in the waveguide structure for modelling this device. In conventional high-power microwave, specific bend structures are used for mode conversion [19]-[21], but in the subwavelength domain, we are the first to report such a design for mode conversion to the best of our knowledge. Section 2 discuss the theory of coupling between the straight waveguide modes, solve the coupled-mode equations for the design proposed in detail, and show the results obtained using MATLAB. Section 3 discusses the results obtained from the device's FDTD simulation and report the net transmission and insertion loss obtained for both the converters. Section 4 summarises the conclusion drawn out of this article.

## 2   Theory and Modelling:

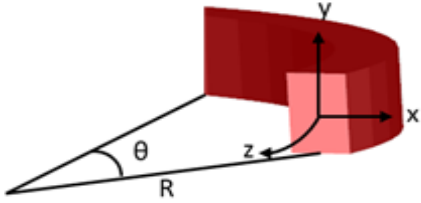

Figure 1 Rectilinear coordinates used to model the device, here R is the radius of bend and θ is the azimuthal angle.

The device proposed here is shown in Figure S1 and consist of two elongated bends at the beginning and the end of the device and a compact arc bend of 25° in the centre. The modelling is done using curvilinear coordinates x, y, and z, as shown in Figure 1. Here, the distance travelled in the azimuthal direction and is given by z=Rθ. Bent waveguide modes can be expanded as a linear combination of the straight waveguide modes [16]-[17]. As the bend modes pass through the bent structure, no coupling between the bent waveguide modes occurs due to the orthogonality property of the modes, but the straight waveguide modes face a refractive index perturbation due to the presence of the bent structure and thus coupling between the straight waveguide modes occurs over the length of the bent waveguide. The refractive index profile's perturbation arises from mapping the bent waveguide to a straight waveguide using the conformal transformation technique. As a result, the refractive index profile of the bent waveguide is no longer a step function [12]-[15]; instead, it becomes a function of direction x and the azimuthal angle and is given by:

$$nc(x,\theta) = n(x)\left(1 + \frac{x-R}{R}\cos\theta\right) \quad (1)$$

The refractive index profile of the bent waveguide for θ=0° is shown in Figure 2(a). As a result of this perturbation, the first two modes of straight waveguide start coupling. However, there is coupling with higher modes and radial modes (due to the bent structure), but in our case, the coupling between the first two straight waveguide modes is

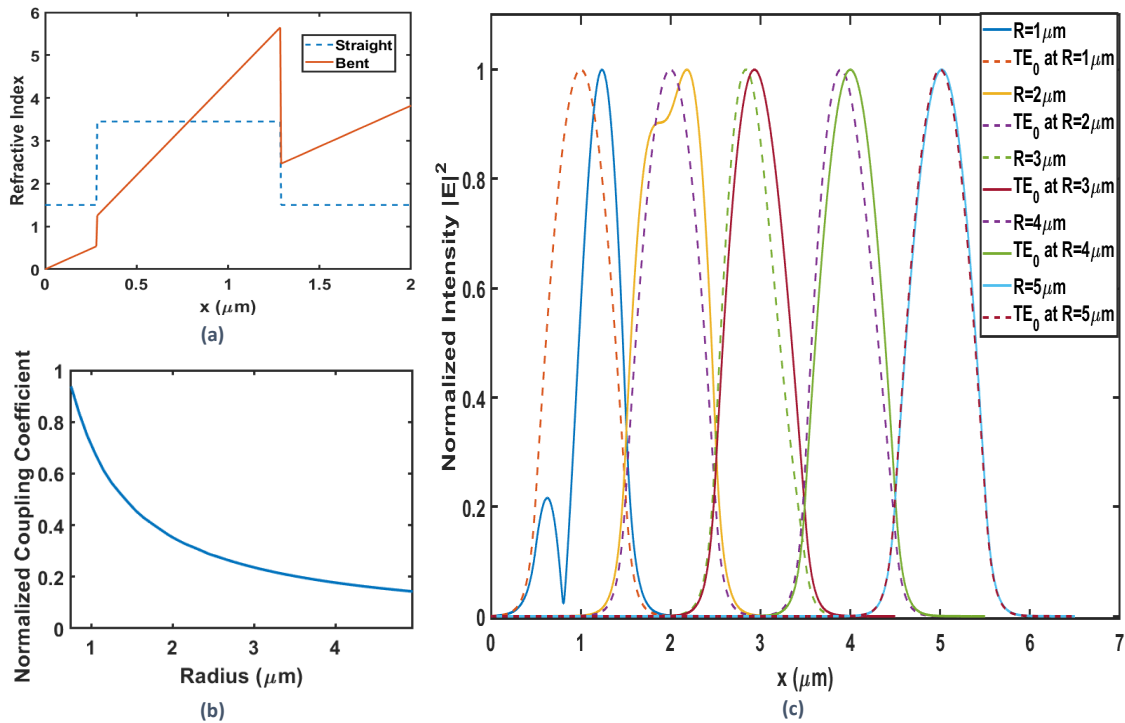

Figure 2 (a) Refractive index profile of straight versus bent waveguide, (b) normalized coupling coefficient versus radius of bend and (c) mode profiles at the output and TE$_0$ mode profile for bends of varying radius.

the most prominent one. The coupling coefficient $\kappa(\theta)$ in this case, given by [18]:

$$\kappa(\theta) = \frac{\omega^2 \mu_o \epsilon_o \iint E_1 \Delta n^2(x,\theta) E_2^* \, dxdy}{\sqrt{\iint E_1 E_1^* \, dxdy} \sqrt{\iint E_2 E_2^* \, dxdy}}, \text{ where } \Delta n^2(x,\theta) = nc^2(x,\theta) - n^2(x) \tag{2}$$

Figure 2(b) shows an inverse relationship between the coupling coefficient and the square of the radius, which agrees with the (2) as $\Delta n^2(x,\theta)$ is inversely proportional to $R^2$. It also interprets that as the bending radius R increases, the coupling between the two straight waveguide modes reduces. Based on (2), we can now write the coupled-mode equations as given by (3).

$$\frac{d}{Rd\theta} A_1 = -i\kappa(\theta) A_2 \tag{3a}$$

$$\frac{d}{Rd\theta} A_2 = -i\kappa(\theta) A_1 \tag{3b}$$

For solving, we substitute (3b) into (3a) and obtain:
$$\frac{d^2}{d^2\theta} A_1 - i\Delta\beta R \frac{d}{d\theta} A_1 + |\kappa(\theta)R|^2 = 0, \quad \Delta\beta = \beta_1 - \beta_2 \tag{3c}$$

Here we observe that $\kappa$ is a function of the direction of propagation, thus on solving the second-order differential equation (3c), we obtain the following transfer matrix and coefficients:

$$\begin{bmatrix} A_1 \\ A_2 \end{bmatrix} = \begin{bmatrix} t_{11} & t_{12} \\ t_{21} & t_{22} \end{bmatrix} \begin{bmatrix} A_2(0) \\ A_1(0) \end{bmatrix} \tag{4a}$$

Here,
$$t_{11} = \exp\left(\frac{iR}{2}\int \Delta\beta d\theta\right)\left[\cos(\int S(\theta)d\theta) - i\Delta\beta R \left(\frac{d\theta}{RdS(\theta)}\right)_{\theta=\theta_{max}} \sin(\int S(\theta)d\theta)\right],$$

$$t_{12} = \exp\left(\frac{iR}{2}\int \Delta\beta d\theta\right)\left[-i\kappa(\theta)R \left(\frac{d\theta}{RdS(\theta)}\right)_{\theta=\theta_{max}} \sin(\int S(\theta)d\theta)\right],$$

$$t_{21} = \exp\left(-\frac{iR}{2}\int \Delta\beta d\theta\right)\left[-i\kappa(\theta)R \left(\frac{d\theta}{RdS(\theta)}\right)_{\theta=\theta_{max}} \sin(\int S(\theta)d\theta)\right] \text{ and} \tag{4b}$$

$$t_{22} = \exp\left(-\frac{iR}{2}\int \Delta\beta d\theta\right)\left[\cos(\int S(\theta)d\theta) - i\Delta\beta R \left(\frac{d\theta}{RdS(\theta)}\right)_{\theta=\theta_{max}} \sin(\int S(\theta)d\theta)\right]$$

Here,
$$S(\theta) = \sqrt{\Delta\beta^2 + +\kappa^2(\theta)}$$

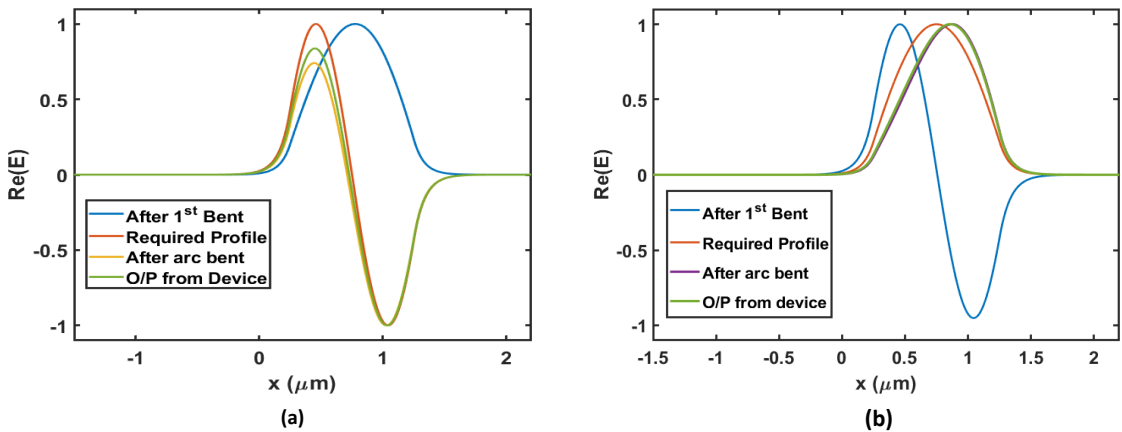

Figure 3 Mode profiles at outputs of different bent sections (a) for $TE_0$ to $TE_1$ converter and (b) for $TE_1$ to $TE_0$ converter.

Based on equations (4a) and (4b), we plotted Figure 2(c) in MATLAB, and the results in Figure 2(c) agree well with Figure 2(b). Figure 2(c) shows that increasing the bend radius reduces the distortion in the mode profile, shifting of the centre and asymmetricity of the mode profile at output; this is caused due to a decrease in coupling coefficient as depicted in Figure 2(b). These results are in agreement with the literature.

We modelled our device with the two elongated bends and a compact arc bend based on the equations (4a) and (4b). The structure begins with an elongated bend, followed by an arc bend. Maximum mode conversion occurs in the central arc bend. This is clearly visible in Figure 4(a) and Figure 5(a). This arc bend is then followed by the elongated bend with a radius of curvature in the opposite direction that enhances the mode transmission in both TE0 to TE1 converter and vice versa by shifting the profile back towards the centre of the waveguide. Due to the high index contrast ratio and the presence of these elongated bends with such a small arc bend, the device losses are minimised. Figures 3(a) and 3(b) show the mode profile at the outputs of the three bends and the required mode profile for $TE_0$ to $TE_1$ and $TE_1$ to $TE_0$ mode converters, respectively. After performing the overlap integral of the output profile with the required profile, we obtained a net transmission of 95.73% and 95.21% for the two mode converters, respectively.

## 3    Simulation and Results:

Based on the above theory, we modelled our devices and performed a simulation using the FDTD module in Lumerical. The structures of the mode converters consist of a 0.22 μm × 1 μm silicon waveguide over silicon dioxide, two elongated bends and an arc bend of radius 0.785 μm. The input signal was given over a bandwidth of 100 nm, and results for mode conversion efficiency/net transmission, insertion loss, and mode profiles at the various stages were obtained. To enhance the net transmission and reduce beating in the case of $TE_1$ to $TE_0$ conversion, a 5 μm long multi-stacked trapezoid (MST) structure [22] was implemented after the device, whose input and output widths are kept the same as that of the waveguide and central widths at four points 1 μm apart were optimised using particle swarm optimisation technique in both the converters.

### 3.1    $TE_0$ - $TE_1$ Mode Converter:

An input signal of $TE_0$ mode with a bandwidth of 100 nm ranging from 1.5 μm to 1.6 μm was given to the structure, as shown in Figure 4(a), which shows the structure's complete field profile at 1.55 μm. As stated in the previous section, the maximum conversion occurs in the compact arc bend, and this is supported by results obtained in Figure 4(a), as we can observe the maximum mode conversion occurs in the arc bend. As the mode passes through the second elongated bend, it becomes more and more like $TE_1$. At the output of this second bend, a mode expansion monitor along with a field monitor was implemented to obtain the mode profile as well as the net transmission of the device; an efficiency of 96.28% was obtained at this point, which is in close agreement with the value of 95.73% obtained with the MATLAB modelling. Figure 4(b) shows the mode profile at this point. After passing through the MST, we again implemented a field monitor to obtain the mode profile shown in Figure 4(c) and net transmission shown in Figure 4(e). Figure 4(d) shows the required mode profile for visual comparison with the converted mode. As Figure 4(e) shows that a net transmission of 98.65% was obtained near the wavelength of 1.55 μm and over the complete 100 nm wavelength range; we observe that net transmission remains above a minimum value of 94.2%. Also, from the results obtained in Figure 4(f), we observe that the insertion loss is very small over the complete operational bandwidth and a minimum value occurs near the wavelength of 1.55 μm. The total length of the device was 10.5 μm; dimensions of MST and a perspective view of the device have been mentioned in Figures S2 and S3.

### 3.2    $TE_1$ to $TE_0$ Mode Converter:

The same structure with different MST dimensions was used for conversion from TE1 to TE0 mode; the device length was again 10.5 μm. Dimensions of MST and a perspective view of the device are displayed in Figures S4 and S5 for the reader's reference. Operational bandwidth remained the same (1.5 μm to 1.6 μm), and TE1 mode was given at the input. As seen in Figure 5(a), we again observed maximum conversion in the arc bend. Monitors were placed at the same positions as in the case of TE0 to TE1 converter. After the second elongated bend, we obtained a net transmission of 95.75% from the mode expansion monitor, which is again in close agreement with the value of 95.21% obtained using MATLAB. Figure 5(b) shows the mode profile at the output of the second elongated bend. After this, the signal was passed through an optimised MST structure for enhancing transmission efficiency. Figure 5(c) and 5(d) shows the mode profile at the output and the required mode profile, respectively; Figure 5(a) verifies that there is no beating of output mode from one wall to another. At the complete device's output, the net transmission of 99.1 % was obtained around the wavelength of 1.55 μm, as shown in Figure 5(e). Figure 5(e) indicates that the net transmission remains above a value of 93.5% for almost the entire operational bandwidth. Results confirm a highly efficient, comprehensive bandwidth operation of the device. Figure 5(f) shows the device's insertion loss over the complete operational wavelength, and it infers that the insertion loss remains minimal over the whole spectrum.

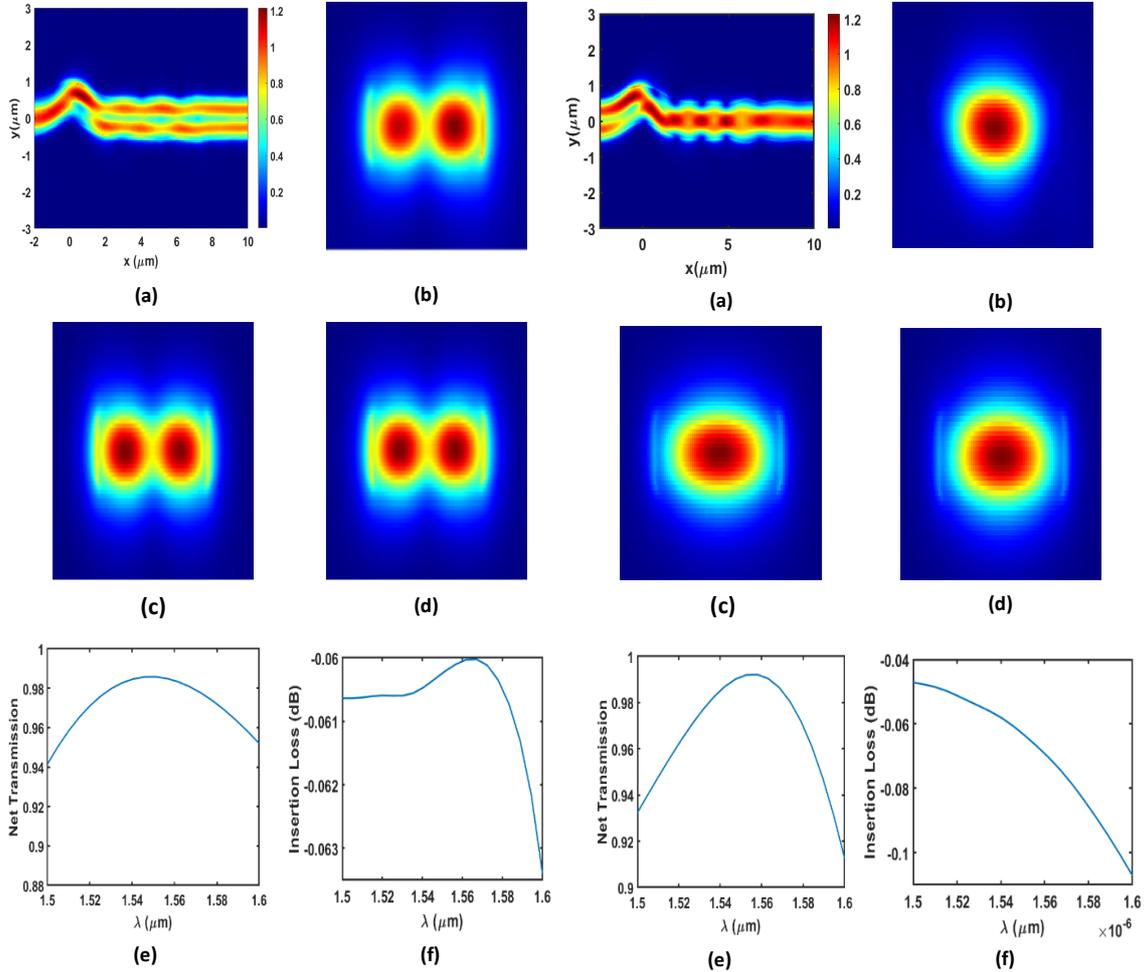

Figure 2 (a) Full profile of the field of device, (b) mode profile at the output of the second elongated bend, (c) output profile at the output of the device, (d) required mode profile, (e) net transmission versus wavelength and (f) insertion loss versus wavelength for $TE_1$ to $TE_0$ Converter.

Figure 5 (a) Full profile of the field of device, (b) mode profile at the output of the second elongated bend, (c) output profile at the output of the device, (d) required mode profile, (e) net transmission versus wavelength and (f) insertion loss versus wavelength for $TE_1$ to $TE_0$ Converter.

## 4  Conclusion:

Concluding the paper, we have theoretically demonstrated compact $TE_0$ to $TE_1$ and $TE_1$ to $TE_0$ mode converters with high efficiency and low losses over a wide operational bandwidth of 100 nm. In both the converters same bent structure was utilised, which fulfils the reciprocity criterion of the optical devices. Results obtained using FDTD simulation 96.28% and 95.75% after the bend structure matches well with the theoretically calculated values of 95.73% and 95.21% for $TE_0$ to $TE_1$ and $TE_1$ to $TE_0$ conversion, respectively. For $TE_0$ to $TE_1$ mode conversion, net transmission of more than 94.2% was obtained over the operational bandwidth of 1.5 μm to 1.6 μm, with the maximum transmission of 98.65% occurring near 1.55 μm. For $TE_1$ to $TE_0$ mode conversion, net transmission of more than 93.5% was obtained over almost the entire operational bandwidth, with a maximum of 99.1% occurring near 1.55 μm. The total device length obtained here is 10.5 μm.

Such compact, highly efficient broadband passive converters can be utilised for on-chip mode conversion in various integrated photonic devices. Due to the bent architecture and compact footprint, these converters can be actively incorporated in the waveguide mesh based programmable photonic integrated circuits (PPIC). In PPIC, the signal passes through multiple tunable switches, making insertion loss an essential factor to be taken into consideration. Low insertion loss, broadband operation and high conversion efficiency in the proposed converters make them an excellent choice for such circuits.

**Acknowledgement:** We would like to acknowledge Mr Prakash Chandra Arya from the Centre for Earth Science of the Indian Institute of Science for his valuable comments and advice for editing the paper. We would also like to extend our thanks to the members of the Applied Photonics Lab of the Electrical Communication Engineering Department at the Indian Institute of Science.

**Disclosure**

The authors declare no conflicts of interest.

**References:**

1. Weimin Ye, Xiaodong Yuan, Yang Gao, and Jianlong Liu, "Design of broadband silicon-waveguide mode-order converter and polarisation rotator with small footprints," Opt. Express 25, 33176-33183 (2017)
2. Frandsen LH, Elesin Y, Frellsen LF, Mitrovic M, Ding Y, Sigmund O, Yvind K., "Topology optimised mode conversion in a photonic crystal waveguide fabricated in silicon-on-insulator material". Opt Express. 2014 Apr 7;22(7):8525-32.
3. D. Ohana, B. Desiatov, N. Mazurski, and U. Levy, "Dielectric metasurface as a platform for spatial mode conversion in nanoscale waveguides," Nano Lett. 16(12), 7956–7961 (2016).
4. Daigao Chen, Xi Xiao, Lei Wang, Yu Yu, Wen Liu, and Qi Yang, "Low-loss and fabrication tolerant silicon mode-order converters based on novel compact tapers". Opt. Express 23, 11152-11159 (2015).
5. Li, C., Liu, D., and Dai, D., "Multimode silicon photonics", Nanophotonics, 8, 227 - 247 (2018).
6. Weimin Ye, Xiaodong Yuan, Yang Gao, and Jianlong Liu, "Design of broadband silicon-waveguide mode-order converter and polarisation rotator with small footprints," Opt. Express 25, 33176-33183 (2017).
7. Li, Z., Kim, MH., Wang, C. et al. "Controlling propagation and coupling of waveguide modes using phase-gradient meta-surfaces". Nature Nanotech 12, 675–683 (2017).
8. Wang, H., Zhang, Y., He, Y., Zhu, Q., Sun, L., Su, Y., "Compact Silicon Waveguide Mode Converter Employing Dielectric Metasurface Structure". Advanced Optical Materials 2019, 7, 1801191.
9. Daniel Pérez, Ivana Gasulla, Prometheus Das Mahapatra, and José Capmany, "Principles, fundamentals, and applications of programmable integrated photonics," Adv. Opt. Photon. 12, 709-786 (2020).
10. Mustapha Remouche, Rabah Mokdad, Ayoub Chakari and Patrick Meyrueis, "Intrinsic integrated optical temperature sensor based on waveguide bend loss", Optics & Laser Technology, Volume 39, Issue 7, (2007).
11. Kim, D.-H.; Jeon, S.-J.; Lee, J.-S.; Hong, S.-H.; Choi, Y.-W. Novel S-Bend Resonator Based on a Multi-Mode Waveguide with Mode Discrimination for a Refractive Index Sensor. Sensors 2019, 19, 3600.
12. Amnon Yariv, Pochi Yeh, "Photonics: optical electronics in modern communications", Oxford University Press, New Your (2007).
13. Ajoy Ghatak, K. Thyagrajan, "Optical Electronics", Cambridge University Press (1991).
14. A. W. Snyder and J. D. Love, "Optical Waveguide Theory". London: Kluwer Academic Publishers (1983).
15. M. K. Chin, S. T. Ho, "Design and Modeling of Waveguide-Coupled Single-Mode Microring Resonators", Journal of Lightwave Technology, VOL. 16, (8), (1998).
16. Andrea Melloni, Federico Carniel, Raffaella Costa, and Mario Martinelli, "Determination of Bend Mode Characteristics in Dielectric Waveguides," J. Lightwave Technol. 19, 571- (2001).
17. Andrea Melloni, Paolo Monguzzi, Raffaella Costa, and Mario Martinelli, "Design of curved waveguides: the matched bend," J. Opt. Soc. Am. A 20, 130-137 (2003).
18. Wei-Ping Huang, "Coupled-mode Theory for Optical Waveguides: An Overview", Journal of Optical society of America A, Vol. 11, (3), (1994).
19. C. W. Yuan, H. H. Zhong, and B. L. Qian, "Design of bend circular waveguides for high-power microwave applications," High Power Laser Particle Beams, vol. 21, no. 2, pp. 255–259, 2009.
20. C. W. Yuan and Q. Zhang, "Design of a – transmission line for high-power microwave applications," IEEE Trans. Plasma Sci., vol. 37, no. 10, pp. 1908–1915, Oct. 2009.
21. H. Li and M. Thumm, "Mode conversion due to curvature in corrugated waveguides," Int. J. Electron., vol. 71, pp.333–347, Aug. 1991.
22. Chunlei Sun, Yu Yu, Guanyu Chen, and Xinliang Zhang, "Ultra-compact bent multimode silicon waveguide with ultralow inter-mode crosstalk," Opt. Lett. 42, 3004-3007 (2017)